 \documentclass[12pt]{article}   
\usepackage{amsmath,amssymb}    
                  
\usepackage{latexsym}
\catcode`@=11
\def\secteqno{\@addtoreset{equation}{section}%
\def\theequation{\thesection.\arabic{equation}}}
\catcode`@=12
\secteqno 
\newcommand{\be}{\begin{equation}}
\newcommand{\ee}{\end{equation}}
\newcommand{\bea}{\begin{eqnarray}}
\newcommand{\eea}{\end{eqnarray}}
\newcommand{\bref}[1]{(\ref{#1})}
\newcommand{\nn}{\nonumber}
  
 \newcommand{\D}{\delta} 
\newcommand{\ep}{\epsilon}

\newcommand{\lam}{\lambda}

\newcommand{\h}{\eta}

 \newcommand{\Lam}{\Lambda}

\newcommand{\ba}{\overline }
\def\6{\partial} \def\7{\tilde} \def\8{\hat}

\def\d{\dot}

\def\pa{\partial}

\def\vs{\vskip 3mm}\def\={{\;=\;}}\def\+{{\;+\;}}

\def\vs{\vskip 4mm}

\def\dif{{\rm d}}
\def\deriv{\@ifnextchar[{\@deriv}{\@deriv[]}}
   \def\@deriv[#1]#2#3{\mathchoice%
{{\dif^{#1}#2\over\dif{#3}^{#1}}}{{\dif^{#1}#2/\dif{#3}^{#1}}}
{{\dif^{#1}#2\over\dif{#3}^{#1}}}{{\dif^{#1}#2/\dif{#3}^{#1}}}}

\def\r2{{r^2}}

\usepackage{color}

\def\d2derpar#1#2{\mathchoice%
{{\partial^2 #1\over\partial #2^2}}%
{{\partial^2 #1/\partial #2^2}}%
{{\partial^2 #1\over\partial #2^2}}%
{{\partial^2 #1/\partial #2^2}}%
}

\def\dif{{\rm d}}
\def\deriv{\@ifnextchar[{\@deriv}{\@deriv[]}}
   \def\@deriv[#1]#2#3{\mathchoice%
{{\dif^{#1}#2\over\dif{#3}^{#1}}}{{\dif^{#1}#2/\dif{#3}^{#1}}}%
{{\dif^{#1}#2\over\dif{#3}^{#1}}}{{\dif^{#1}#2/\dif{#3}^{#1}}}}

\def\srP{\sqrt{P^2}}
\def\bigravity{{ bigravity }}

\begin{document}
\thispagestyle{empty}
{\hfill {\rm CERN-PH-TH/2013-284,\, ICCUB-13-243}}
\vskip 2cm
\begin{center} 
{\Large {Dynamical sectors of a relativistic two particle model}}
\vskip 15mm 
{\large
{Daniele Dominici${}^{a}$},
 {Joaquim Gomis${}^b$\,${}^c$}, \\ \vskip 3mm 
{Kiyoshi Kamimura${}^d$} and 
 {Giorgio Longhi${}^a$}
}\vskip 1cm
{ ${}^a$ 
  Department of Physics and Astronomy, University of Florence, 50019 Sesto F., Firenze, Italy; Sezione di Firenze,
INFN, 50019 Sesto F., Firenze, Italy}

{${}^b${ Theory Group, Physics Department, CERN, CH1211, Geneva 23, Switzerland}}

{${}^c${ Departament d'Estructura i Constituents de la Mat\`eria and Institut de Ci\`encies del Cosmos, 
Universitat de Barcelona, Diagonal 645, 08028 Barcelona, Spain}}

{${}^d$ Department of Physics, Toho University, Funabashi, Chiba 274-8510, Japan}

%
\end{center}
\vskip 3cm

\abstract
We reconsider a model of two relativistic particles interacting via a multiplicative potential, 
as an example of  a  simple dynamical system with sectors, or branches,  with different dynamics and degrees of freedom.
The presence or absence of sectors depends on the values of rest masses.
 Some aspects of the canonical  quantization are described.
The model could be interpreted as a bigravity model  in one dimension.

\vskip 10mm

\eject


\section{Introduction \label{sec:0}}
\vs

Usually a physical system has a definite number of physical degrees of freedom. 
However it sometimes happens that it has {\it sectors}, i.e.,   branches of phase spaces
that have different canonical structures, such as different physical degrees of freedom, 
or gauge symmetries.
Generically,  these sectors appear for dynamical systems when the rank of the matrix of Poisson brackets of primary constraints is not constant,
in other words when
the regularity conditions are not verified, see for example \cite{Henneaux:1992ig}.
In these cases there  are non-unique solutions of the stabilization condition in
 the Dirac's  algorithm of constraints.
{ The presence of sectors 
was also analyzed in other contexts, for example when the Legendre transformation
and the
Hamiltonian are multivalued \cite{Henneaux:1987zz}  \cite{Shapere:2012nq} \cite{Wilczek:2012jt}
\cite{Curtright:2013nya}.}

A model of the Zwei-Dreibein gravity has been proposed in a very recent
paper \cite{Bergshoeff:2013xma}. The presence or absence of the Boulware-Deser 
mode in the model has been reanalyzed in \cite{Banados:2013fda}; 
the Hamiltonian analysis  shows in general the existence of sectors, 
and, in each sector, the system has   different number of degrees of freedom. 
For a suitable choice of the parameters
there appears only one sector where the Boulware-Deser mode is absent\footnote{ 
We acknowledge Eric Bergshoeff and Paul Townsend for discussions on this point.}.
In the case of  four dimensions,  bigravity contains also sectors, see for example 
\cite{Alexandrov:2013rxa}.
{ For other models with sectors see  \cite{Lusanna:1991je}
\cite{Banados:1995mq}
 \cite{Saavedra:2000wk} 
\cite{Ivanov:2005vh}.}

In order to have
 a better  understanding of how to deal with  dynamical systems with 
sectors  at classical and  quantum 
 levels, we consider a simple model 
of two relativistic particles  \cite{Kamimura:1977vi,
Kamimura:1977dv, 
Dominici:1977fh, 
Dominici:1978yu}. 
The model is described by a sum of two relativistic particle Lagrangians in which 
the interaction is introduced by replacing their rest masses with 
potentials that depend on the Minkowskian distance of the two particles\footnote{Here the metric is $(+;---)$}, 
\be\label{oldDGLK} 
L=-\sqrt{(m_{10}^2-V({\r2}))\,{{\dot x}_1}^2}-
     \sqrt{(m_{20}^2-V({\r2}))\,{{\dot x}_2}^2}
=-\sum_{j=1,2}  \sqrt{m_j^2({\r2})\,{{\dot x}_j}^2},
\ee 
where ${x_j}(\tau), (j=1,2)$ are the space-time coordinates of the two particles.  
$V({\r2})$ is any  Poincar\'e invariant function of the squared relative distance  
 $r^2=(x_2-x_1)^2$.
$m_{j0}$'s are the rest masses of the particles and 
$m_{j}^2({\r2})=m_{j0}^2-V({\r2})$ are the effective masses of the particles. 
The interaction breaks the individual invariance under diffeomorphism (Diff)
 of the action of two free particles, leaving a universal Diff invariance.

 The canonical action is given by
 \be
 S=\int d\tau \sum_{j=1,2}(p_i\dot x_i-\frac {e_i}{2}( p^2_j-m_j^2({\r2}))=\int d\tau \sum_{j=1,2}(p_i\dot x_i-{e_i}\phi_i),
 \ee
 where $e_i$ are Lagrangian multipliers  that we can interpret as two einbein variables, and  
$\phi_j ={1/2}(p^2_j-m_j^2({\r2}))$ are the mass-shell constraints.

The matrix of the Poisson brackets of the primary constraints 
 \bea
 \begin{pmatrix} \{\phi_1,\phi_1\}& \{\phi_1,\phi_2\}\cr  \{\phi_2,\phi_1\}& \{\phi_2,\phi_2\}
 \end{pmatrix}=
\begin{pmatrix} 0&V'({\r2}) (P \cdot r)\cr-V'({\r2}) (P \cdot r)&0
 \end{pmatrix},
 \eea
where 
$P=p_1+p_2$ is the total { momenta} of the system and  $V'({\r2})=\frac d{dr^2} V({\r2})$, and
has not constant rank. The rank is two if $V'({\r2}) (P \cdot r)\neq 0$ and it is 0 if
$V'({\r2}) (P \cdot r)=0$,  therefore the models has sectors\footnote{
The presence of sectors was already  noticed in 
\cite{Kamimura:1977vi,
Kamimura:1977dv, 
Dominici:1977fh, 
Dominici:1978yu} but only one sector was studied in detail.}. The detailed structure of the sectors will be studied in the next section for the case of the harmonic oscillator.

 If we eliminate the momenta $p_i$ by using their equations of motion we get
\be\label{DGLKeinbein0}
L=\sum_{j=1,2}\left(\frac{\dot x_j^2}{2e_j}+\frac{e_j}{2}m_j^2({\r2})\right).\qquad 
\ee
In this form the Lagrangian can be reinterpreted as the Lagrangian of a model of 
bigravity in one dimension.
Note that for this Lagrangian the primary constraints $\phi_i$ of  \bref{oldDGLK} 
appear as secondary constraints. If we restrict the $V({\r2})$  
to an harmonic  potential, it can be shown that, in the case of equal rest masses ($m_{10}=m_{20}\equiv m_{0}$), the model has a new extra Noether gauge symmetry,  
in addition to Diff The existence  of two gauge symmetries depending on 
the phase space regions, $P^2=0$ or $P^2>0$,  
is a consequence of the fact that the model has sectors,
with different degrees of freedom in each sector. 

In this note we study in detail the appearance of { sectors} at Hamiltonian level. 
As we will see for unequal rest masses there is only one sector. It is a massive sector in the sense that $P^2\neq 0$. This sector contains first and second class constraints. 
The physical degrees of freedom are those of a system of two massive particles.
If the rest masses are equal, we have three sectors: one with
$P^2\neq 0$ like in the previous case and two sectors with $P^2=0$ with 
different number of first and second class constraints.
Although the massless sectors are empty (no classical solution) 
for positive rest mass square ($m^2_{0}>0$),
they are not so for tachyonic rest masses($m^2_{0}<0$).

We  perform the canonical quantization of the two sectors. 
For the massive sector it is useful to 
consider first  a canonical transformation at classical level \cite{Kamimura:1977dv}  \cite{Dominici:1978yu} such that the second class constraints become a pair of canonical
variables \cite{Shanmugadhasan:1973ad} \cite{Castellani:1978yv}.
This allows to impose the second class constraints on the physical states,  by considering a non-hermitean combination of them.  We get a spectrum of increasing 
masses for higher internal spins.  In the case of unequal rest masses there is a branch of the mass spectrum which contains ghosts.
Instead, for the massless sector, there are no  physical states for physical  particles with positive rest mass  square corresponding to the fact that there are no classical solutions.
If we consider tachyons this sector is not empty, there are states with 
 helicities depending on the value of the tachyonic mass. 

The organization of this paper is as follows: in Section 2   we introduce the model and perform the Hamiltonian analysis. In Section 3 we give the gauge transformations. Section 4 is devoted to the canonical quantization and finally in Section 5 we give some conclusions and an outlook.

\section{ Sectors of an interacting  relativistic two particle model}

We reconsider a model of two relativistic  interacting particles via a multiplicative potential 
introduced in 
\cite{Kamimura:1977vi,
Kamimura:1977dv, 
Dominici:1977fh, 
Dominici:1978yu}.
The Lagrangian is  given in \bref{oldDGLK}.
We rewrite the Lagrangian by introducing two einbein variables ${e_j}$, see for example \cite{carles}  \cite{Kalashnikova:1996pu},
\be\label{DGLKeinbein}
L=\sum_{j=1,2}\left(\frac{\dot x_j^2}{2e_j}+\frac{e_j}{2}m_j^2({\r2})\right).\qquad 
\ee
 In order to have a well defined Lagrangian we will assume $e_j\neq 0$. 
In addition $e_1,e_2$ should have the same sign in order to reproduce \bref{oldDGLK}, so
$e_1+e_2\neq 0$. In this form the model can be reinterpreted as a \bigravity in one dimension.
 In this paper we will consider a special case of harmonic potential 
$m_j^2({\r2})=m^2_{j0}-\kappa^2 r^2$
with the parameters, the rest masses $m^2_{j0}$ and  $\kappa$.
The Lagrange equations of motion are
 \bea\label{eomx}
 \left[L\right]_{x_j}&=&\frac{\pa L}{\pa x_j}-\frac{d}{d\tau}\frac{\pa L}{\pa\dot x_j}=(-)^{j+1}\kappa^2(e_1+e_2)r
 +\frac{\dot e_j\dot x_j}{e_j^2}-\frac{\ddot x_j}{e_j}=0,
 \\
 \left[L\right]_{e_j}&=&\frac{\pa L}{\pa e_j}-\frac{d}{d\tau}\frac{\pa L}{\pa\dot e_j}=\frac 12 (\frac{\dot x^2}{e_j^2}-m^2_j)=0.
 \eea

In the Hamiltonian formalism the canonical momenta are
\be
p_j=\frac{\pa L}{\pa\dot x_j}=\frac{\dot x_j}{e_j},\qquad 
\pi_j=\frac{\pa L}{\pa\dot e_j}=0
\ee
and we have two primary constraints
\be
\pi_j=0,\qquad (j=1,2).
\ee
The canonical Hamiltonian is given by 
\be
H_c= {e_j}\,\phi_j,\qquad
\phi_j\equiv\frac12(p_j^2-m_j^2({\r2})),
\ee
and the Dirac Hamiltonian, that includes the primary constraints, is
\be
H=H_c\,+\pi_j\Lam_j,
\ee
where $\Lambda_j$'s are arbitrary functions of $\tau$ that  work as Lagrangian multipliers.
The Hamilton equations of motion are
\be\label{eom}
\dot e_j=\Lam_j,\qquad \dot x_j=e_jp_j,\qquad 
\dot \pi_j=-\phi_j,\qquad \dot p_j=(-1)^{j+1}  (e_1+e_2)\kappa^2\, r,
\ee
which imply that the total momentum $P=p_1+p_2$ is conserved.

The  Dirac's  algorithm starts to examine stability condition of the primary constraints
$\pi_j=0 $. It produces the secondary constraints 
\be
 \phi_j=0 
\ee 
and the stability of these constraints gives
\be
\dot\phi_1= \kappa^2(P\cdot r)\,e_2=0 ,\qquad
\dot\phi_2=- \kappa^2(P\cdot r)\,e_1=0.
\ee
{Note that the matrix Poisson brackets of the secondary constraints has no constant rank, in fact
$\{\phi_1,\phi_2\}= \kappa^2 (P\cdot r)$.}

 Since $e_j$'s are not vanishing by initial hypothesis of
 the Lagrangian \bref{DGLKeinbein}, in addition  
 $e_1=e_2=0$ only gives  trivial system $H\equiv 0$, 
we have a tertiary constraint
\be
\Theta\equiv(P\cdot r)=0.
\ee
 Now we compute the Poisson bracket of $\Theta$ with $\phi_i$ 
\bea
\begin{pmatrix} \{\Theta,\phi_1\},& \{\Theta,\phi_2\}
\end{pmatrix}=\begin{pmatrix}  -\frac 12 P^2-\frac 12 (m_{10}^2-m_{20}^2), &-\frac 12 P^2+\frac 12 (m_{10}^2-m_{20}^2) 
\end{pmatrix},
\eea
 where 
the constraints $\phi_i=0$ have been used.
 The above rectangular matrix has not constants rank. It has rank 0 for $P^2=m_{10}^{2}-m_{20}^2=0$ and rank  1 for others
, therefore the model has further ramifications.

The stability of the tertiary constraint  requires  the relation 
 \bea
\chi
 &=& \dot\Theta=P\cdot(e_2p_2-e_1p_1)
 \nn\\&=&-
\frac{(e_1-e_2)}2P^2-(e_1+e_2)(\phi_1-\phi_2+\frac12(m_{1}^2(r^2)-m_{2}^2(r^2)))
\nn\\&=& 
-\frac12({e_-}\,P^2+{e_+}\,\Delta_- )
=0, 
\label{chiiii} \eea
where we have used the secondary constraints $\phi_j=0$ and defined
\be
e_\pm=e_1\pm e_2,\,\,\,  \Delta_- =(m_{10}^2-m_{20}^2).
\ee

Now we study the different sectors of this constraint depending on 
the values of  $\Delta_-$.
For the model with different rest masses  $\Delta_- 
\neq0$, since $e_+\neq0$, \bref{chiiii} implies $P^2\neq0$. Solving it for
 $e_-$,  the quaternary constraint takes the form
\bea
&& e_-+\frac{\Delta_-}{P^2}\,e_+=0.   \qquad  (massive \; sector) 
\label{sect4}\eea
 The stability of this constraint gives a relation between the arbitrary functions,
\be
\Lambda_-+\frac{\Delta_-}{P^2}\,\Lambda _+=0.
\ee
There is only one sector for  $\Delta_- =(m_{10}^2-m_{20}^2)\neq0$.
The first class combinations of constraints are 
\be
\pi_+-\frac{\Delta_-}{P^2}\pi_-=0,\qquad 
\phi_+-\frac{\Delta_-}{P^2}\phi_-=0,
\label{4.10}\ee 
with $\phi_\pm=\phi_1\pm\phi_2$, and 
\be
\pi_-=0,\quad e_-+\frac{\Delta_-}{P^2}\,e_+=0,\quad 
\phi_-=-(P\cdot q)-\frac{ \Delta_-} 2=0,\quad \Theta=(P\cdot r)=0
\label{4.1}\ee
are the second class  constraints, where $q=\frac 12 (p_2-p_1)$ is the relative momentum.

\vs
For the case of equal  rest masses  $\Delta_- =(m_{10}^2-m_{20}^2)=0$, \bref{chiiii}
requires either  
\bea
case\;1,&\quad& e_-=0, \quad P^2\neq 0, \quad {\rm then} \quad \Lambda_-=0,\quad  {\rm (massive \; sector),} 
\\ \nn\\
case\;2,&\quad& P^2=0, \quad e_-\neq 0, \qquad {\rm(no\;more\;cond.\;: massless  \; sector),}
\label{mlsec}
\\ \nn\\
case\;3,&\quad& P^2=0\quad{\rm and}\quad  e_-=0, \quad{\rm then}\quad \Lambda_-=0.\quad {\rm (massless \; sector).} 
\label{mlsec3}\eea

\vs
The {\it case }1 is just $\Delta_-=0$ of previous case \bref{sect4}.
The constraints appearing here 
\be
\pi_+=0,\qquad
\phi_+=0,
\ee
are first class while 
\be 
\pi_-=0,\quad e_-=0,\qquad \phi_-=-(P\cdot q)=0,\quad (P\cdot r)=0
\ee
are second class. The constraint  $\phi_+=0$ determines the mass of the system, while the last two kill the longitudinal relative coordinate $r^\mu$ and momentum $q^\mu$.
The physical degrees of freedom are given by
\be
2\times 2+ 8\times 2-2\times 2-4=12,
\ee
which is the physical dimension of a system of two massive particles.
\vs

In the {\it case} 2 all the constraints 
\be
\pi_\pm=0,\quad \phi_\pm=0,\quad (P\cdot r)=0,\quad P^2=0
\ee
are first class and 
the dimension of the physical phase space  is
\be
2\times 2+ 8\times 2-2\times 6=8,
\label{massless dim}\ee
which is the physical phase space of a system with two massless particles.
 
  In the {\it case} 3 the constraints 
\be
\pi_+=0,\quad \phi_\pm=0,\quad (P\cdot r)=0,\quad P^2=0,
\ee
are first class and
\be
\pi_-=0,\qquad e_-=0
\ee
are second class. 
In this case the dimension of the physical phase space  is
\be
2\times 2+ 8\times 2 -2\times 5-2=8,
\label{massless dim3}\ee
which is same as in the (above) case 2. 
Despite the fact that counting of degrees of freedom is eight, 
as we will see in the next section, the constraints have no solution for positive rest mass
case, $m_0^2>0$, thus these sectors are empty. 

Now we would like to see which is the evolution of the Lagrangian multipliers in the different sectors. If $e_-=0$ at $\tau =0$,  integrating \bref{eom} the evolution in $\tau $ is given by
  \be
  e_-=\int^\tau_0 \Lambda_-(\tau') d\tau
\ee
which, for generic $\Lambda_-$,  gives $e_-\neq 0$. However since in the case 1 and 3 $\Lambda_-=0$, 
 $e_-$ will remain zero during the evolution  of the system. Instead, if we are in  case 2,  
$P^2=0, \;e_-\neq 0$. If $\, e_-(0)=a$  at $\tau =0,\,$ the evolution in this case is given
by
\be
e_-=\int^\tau_0 \Lambda_-(\tau') d\tau'+a.
\ee
We always can choose $\Lambda_-$ such that at given $\tau$ we have $e_-(\tau)=0$ and we are no
longer in case 2.
{However we are not in case 3, since in this case too $\Lambda_-=0$.
The system will evolve and it could  move again to $e_-\neq 0$.}
It seems that from the point of view of the true degrees of freedom the evolution among sectors does not matter since the space-time variables are unaffected by this phenomenon.
Evolution among sectors occurs in the model studied in \cite{Saavedra:2000wk}\footnote{{We acknowledge discussions with Jorge Zanelli in this point.}}.

\section{Gauge symmetries}

Let us now study the gauge symmetries in these  sectors. 
In the massive sector with equal or different rest masses we have only one gauge transformation in agreement with the presence of only one primary first class constraint.
The transformation is given by
\bea
\D e_j&=&\frac{d}{d\tau}(\ep e_j),
\qquad
\D x_j=(\ep e_j)p_j=\ep \dot x_j
\label{Difftr}\eea
where $\ep=\ep(\tau)$ is an arbitrary function. It is a Noether symmetry since
\bea
\D L= \frac{\dot x_j}{e_j} \frac{d}{d\tau}(\ep \dot x_j)-
\frac{\dot x_j^2}{2e_j^2} \frac{d}{d\tau}(\ep e_j)+\frac{m^2}2 \frac{d}{d\tau}(\ep e_j)
-e_j\kappa^2\,r(\ep\dot r)= \frac{d}{d\tau}(\ep L).
\eea
It is the well known diffeomorphism (Diff) transformation which has  a closed algebra
\be
[\D_{\ep_2},\D_{\ep_1}]e_j=\D_{\ep_3}e_j,\qquad
[\D_{\ep_2},\D_{\ep_1}]x_j=\D_{\ep_3}x_j,\qquad
\ep_3=\ep_1\dot\ep_2-\ep_2\dot\ep_1.
\ee

For the massless sector \bref{mlsec}, with equal masses,  we have two independent primary first class constraints. This is a signal that we will have two gauge transformations, one being the Diff. as in the massive case \bref{Difftr} and the other one a new transformation with gauge parameter $\lambda(\tau)$,
\bea
\D e_j&=&
(-)^{j+1}\left[
\frac{d}{d\tau}\left\{\frac{1}{\kappa^2(e_1+e_2)}\frac{d}{d\tau}
\left(\frac{-1}{(e_1-e_2)}{\dot\lam}\right)\right\}
-\frac{(e_1+e_2)}{(e_1-e_2)}{\dot\lam}\right],
\nn\\
\D x_j 
&=&(-)^{j+1}\,\left\{\frac{1}{\kappa^2(e_1+e_2)}\frac{d}{d\tau}
\left(\frac{-1}{(e_1-e_2)}{\dot\lam}\right)\right\}\frac{\dot x_j}{e_j}+
\frac{ 1}{(e_1-e_2)}{\dot\lam}\,r+\lam \,(\frac{\dot x_1}{e_1}+\frac{\dot x_2}{e_2}),
\nn\\ \label{MLtr}\eea
which is also a Noether symmetry, indeed,
\bea
\D L 
&=& \frac{d}{d\tau}\left[
\frac{1}2\left\{\frac{1}{\kappa^2(e_1+e_2)}\frac{d}{d\tau}
\left(\frac{-1}{(e_1-e_2)}{\dot\lam}\right)\right\}
\{(\frac{\dot x_1}{e_1})^2-(\frac{\dot x_2}{e_2})^2\}+
\frac{\lam}2\,(\frac{\dot x_1}{e_1}+\frac{\dot x_2}{e_2})^2\right].
\nn\\ \eea

In this case the algebra of gauge transformation closes only on shell as 
\bea
[\D_{\lam_2},\D_{\lam_1}]e_i=\D_{\lam_3}e_i,\qquad
[\D_{\lam_2},\D_{\lam_1}]x_i=\D_{\lam_3}x_i-\frac{\ep_{ij}[L]_{x_j}}{k^2e_1e_2} \,\frac{d}{d\tau}\left(\frac{(\lam_1\dot\lam_2-\lam_2\dot\lam_1)}{(e_1-e_2)}\right), 
\eea
where  $[L]_{x}$ are  the equations of motion for $x_j$ \,\bref{eomx} and 
\bea
\lam_3=\frac{2}{k^2(e_1+e_2)(e_1-e_2)^2}(\dot\lam_1\ddot\lam_2-\dot\lam_2\ddot\lam_1).
\eea

 The second gauge transformation $(\D_{\lam})$  and diffeomorphism transformation $(\D_{\ep})$  close off shell as  
\be
[\D_{\ep},\D_{\lam}]e_j=\D_{\7\lam}e_j,\qquad
[\D_{\ep},\D_{\lam}]x_j=\D_{\7\lam}x_j,\qquad
\7\lam=-\ep\dot\lam.
\ee
Note that the second gauge transformation exists only when $e_\pm$ are not vanishing.
The gauge structure of a theory is encoded in the BV formalism \cite{Batalin:1981jr}. For a review see for example, \cite{Henneaux:1989jq}   \cite{Gomis:1994he}. The construction of the classical master equation and the BRST symmetry will be given elsewhere 
\cite{progress}.

\section{Canonical quantization}

Here we will discuss the canonical quantization of the model for the massive and massless sectors.
\subsection{Massive Sector}
The quantization of the massive sector of the model for unequal   rest masses ($\Delta_-\neq0$),  and also for equal mass case ($\Delta_-=0$),  is performed following 
\cite{Dominici:1978yu}. The constraints  in \bref{4.10}-\bref{4.1} are  
\bea
&& \pi_+-\frac{\Delta_-}{P^2}\pi_-=0,\qquad 
\phi_+-\frac{\Delta_-}{P^2}\phi_-=0,\qquad (1cl.) 
\label{4.11cl}\\
&&\pi_-=0,\quad e_-+\frac{\Delta_-}{P^2}\,e_+=0,\quad 
\phi_-=-(P\cdot q)-\frac{ \Delta_-} 2=0,\quad \Theta=(P\cdot r)=0,\quad (2cl.). 
\nn\\ \label{4.12cl}\eea
Since the latter four are the second class constraints, we first perform a canonical transformation such that they become  new sets of canonical pairs.
\footnote{{
The theorem which guarantees the existence of this the canonical transformation is given by \cite{Shanmugadhasan:1973ad,Castellani:1978yv}.}}
It is generated by
 \be
 W(\pi,P,q,\bar e,\bar x,u) =\frac12\bar e_+( \pi_+-\frac{\Delta_-}{P^2}\pi_-) +\frac12\bar e_-\pi_-+
\bar x^\mu P_\mu
+ u^{(\lambda)}\, {\ep_{(\lambda)}}^\mu(P)\,q_\mu+u^{(0)}\, \frac{\Delta_-}{2\srP},
 \ee
 where the polarization vectors $ {\ep_{(\lambda)}}^\mu,\,(\lam=0,1,2,3)$ are 
 \bea
  {\ep_{(\lambda)}}^\mu=
\begin{pmatrix} {\ep_{(0)}}^\mu \cr {\ep_{(\lam')}}^\mu  \end{pmatrix}
=\begin{pmatrix}
\frac{P^0}{\srP}&\frac{P^j}{\srP}\cr
\frac{P_{\lam'}}{\srP}&{\delta_{\lam'}}^j-\frac{P_{\lam'} P^j}{\srP(P_0+\srP)}
 \end{pmatrix},
\eea
and $\lam'=1,2,3,\, j=1,2,3.$
The new canonical variables $(\bar e_i,\bar\pi_i,\bar x^\mu,\bar P_\mu,u^{(\lam)},v_{(\lam)})$ are 
related to the old ones  $(e_i,\pi_i, x^\mu=\frac 12(x^\mu_1+x^\mu_2), P_\mu,r^\mu,q_\mu)$ by 
\cite{Dominici:1978yu}
 \bea
\bar \pi_+&=&\frac{\pa W}{\pa\bar e_+}=( \pi_+-\frac{\Delta_-}{P^2}\pi_-), \quad \bar\pi_-=
\frac{\pa W}{\pa\bar e_-}=
\pi_-, \quad 
e_+=\bar e_+, \quad  e_-= \bar e_--\bar e_+\frac{\Delta_-}{P^2},
\nn\\
x^\mu&=& \frac{\pa W}{\pa P_\mu}= 
\bar x^\mu +
u^{(\lambda)}\,\frac{\pa {\ep_{(\lambda)}}^\nu}{\pa P_\mu}\,q_\nu-
\frac{\Delta_-u^{(0)}}{2\srP^3}P^\mu+\bar e_+ \frac{2\Delta_-}{(P^2)^2}\pi_-P_\mu,\quad
\ba P_\mu=\frac{\pa W}{\pa \bar x^\mu}=P_\mu,
\nn\\
r^\mu &=& \frac{\pa W}{\pa q_\mu}=
u^{(\lambda)}\, {\ep_{(\lambda)}}^\mu ,\quad
v_{(\lambda)} =\frac{\pa W}{\pa u^{(\lambda)}}
= {\ep_{(\lambda)}}^\mu\,q_\mu+ 
\frac{\Delta_-}{2\srP}{\delta_\lam}^0.
\eea
The second class constraints in \bref{4.12cl} are 
\bea
&&\bar\pi_-=0,\quad \bar e_-=0,\quad 
\phi_-=-\srP\, v_{(0)}=0,\quad \Theta=\srP \, u^{(0)}=0 \label{secondclass}
\eea
and the first class constraints in \bref{4.11cl} are 
\bea
\bar\pi_+=0,\quad  \frac14C_0\equiv
\phi_+-\frac{\Delta_-}{P^2}\phi_-=\frac14P^2-\sum_{\lam'=1}^3(v_{(\lam')}^2+\kappa^2 {u^{(\lam')}}^2)
-\frac{\Delta_+}2+\frac{\Delta_-^2}{4P^2}=0,
\label{firstclass}
\eea
with $\Delta _+=(m_{10}^2+m_{20}^2)$. Notice that in $C_0$ we have deleted  terms of square of constraints. 
 They are the first class constraints which commute strongly with the second class constraints. This result is similar to what happens when using Dirac brackets: we can put to zero second class constraints  in the expression of the first class ones.  However the canonical transformation method allows to work in the complete phase space.

The canonical quantization is obtained  by imposing the commutation relations on the  canonical pairs 
\be
[\bar e_i, \bar \pi_j ] =- i\D_{ij}, \quad 
[\bar x^\mu,P_\nu]=-i {\D^\mu}_\nu,\;\,\,\,[u^{(\lambda)},v_{(\rho)}]=-i{\delta^{\lambda} }_
{\rho}, 
\ee
and requiring the first class constraint (\ref{firstclass}) as operatorial condition on the physical states. Concerning the second class constraints,   
 a non-hermitian combination of eqs.(\ref{secondclass}),  
following Gupta-Bleuler,
is required so that
we have
\be
\frac 1 {\sqrt{2}} (
\bar e_--i\bar\pi_-)|\psi_{phys}\rangle=0,\qquad a_{(0)} |\psi_{phys}\rangle =0 ,
\ee
where we have defined the  operators $a_{(\lambda)}$, for $\lambda=0,1,2,3$, as
\be
a_{(\lambda)}=\frac 1 {\sqrt{2{\kappa}}}(v_{(\lambda)}- i {\kappa} u_{(\lambda)}    ) 
\ee
such that
\be  
[a_{(\lambda)},a^\dagger_{(\rho)}]=-g_{\lam\rho}.
\ee
These two conditions imply
\be
\langle\psi_{phys}\vert \bar e_- \vert \psi_{phys}\rangle =
\langle\psi_{phys}\vert \bar \pi_- \vert \psi_{phys}\rangle =
\langle\psi_{phys}\vert u^{(0)} \vert \psi_{phys}\rangle =\langle\psi_{phys}\vert v_{(0)} \vert \psi_{phys}\rangle =0
\ee
and are satisfied, in the coordinate representation, by 
 a gaussian dependence of the wave function on  the variables $\bar e_-$ and $u_{(0)}$.

Finally we require the first class constraints \bref{firstclass} as the  physical state condition,
\be
\bar\pi_+|\psi_{phys}\rangle =0,
\ee
implying  that the wave function, in the coordinate representation, does not depend on $\bar e_+$ and
 \be
C_0 |\psi_{phys}\rangle =0, 
 \label{PSC}\ee
 where 
\be
C_0=P^2-4\kappa\sum_{\lam'=1}^3(a^\dagger_{(\lambda')}a_{(\lambda')}+a_{(\lambda')}a^\dagger_{(\lambda')})-2 \Delta_+ +\frac {\Delta_-^2} {P^2}.
\label{C0b}
\ee
Notice that for unequal masses,  the propagator  has the general form
\be
\frac 1 {M_1^2-M_2^2} [\frac 1 {p^2-M_1^2}-\frac 1 {p^2-M_2^2}],
\ee
where $M_1^2$ and $M_2^2$ are two solutions in $P^2$ of the primary constraint (\ref{PSC}).
Therefore due to the minus sign in the second term the theory in general contains ghosts.
The Hilbert space would be the direct sum  of two Hilbert spaces one with positive norm and the second one with negative norm. As soon as one add  interaction terms in the field theory Lagrangian, instabilities can occur and special care has to be taken
\cite{Hawking:2001yt}.

For this reason  we limit  the study of  the spectrum in the simplest case of equal masses $\Delta_-=0$
so that the wave equation becomes quadratic in the momentum.
Using \bref{PSC} the mass operator ${\cal M}^2$ is given by
\be
{\cal M}^2 
=8{\kappa} [\sum_{\lam'=1}^3 a_{(\lambda')}^\dagger a_{(\lambda')}+\frac 3 2 +\frac {m_0^2} {2{\kappa}}].
\ee
Standard procedure allows to built the quantum states of the internal space corresponding to a three dimensional harmonic oscillator. For increasing occupation numbers one can build states of increasing internal angular momentum which can be given by the representation of 
O(3). 

The Lorentz generators 
\be
M_{\mu\nu}=x_\mu p_\nu-x_\nu p_\mu+r_\mu q_\nu-r_\nu q_\mu
\label{angmom} 
\ee
expressed in terms of the new canonical variables 
are giving the massive representation,  
\bea
M_{ij}&=& \bar x_i P_j-\bar x_j P_i+
T_{ij}
\nn\\
M_{0i}&=&\bar x_0 P_i-\bar x_i P_0+ \frac {
 P^{j} }{P^0+\sqrt{|{\bf P}|^2}}T_{ji},
\eea
where $T_{\rho'\lambda'}$ is the O(3) little group generator,
\be
T_{\rho'\lambda'}=u_{(\rho')}v_{(\lambda')}-u_{(\lambda')} v_{(\rho')}.
\label{spin0}
\ee


  \subsection{Massless sector}

 Let us now consider the quantization of the  sectors with $P^2=0$.
  In both sectors for the space-time
 variables all constraints are first class and given by
\bea
&&\pi_i=0,\,\,P^2=0,\qquad \label{mass}\\
&&q^2+k^2r^2-m_0^2=0,\label{relat}\\
&&(P\cdot  r)=0,\,\quad 
(P\cdot  q)=0\label{trans}
\eea
or the equivalent combinations
\be  \pi_\pm=0,\quad 
\phi_+=0,\quad
\phi_-=-(P\cdot q)=0,\quad \Theta=(P\cdot r)=0, \quad 
\chi=\frac12P^2=0.  
 \label{4.22cl}\ee
The construction of the Hilbert space can be performed by following the covariant quantization of the electromagnetic field,
 introducing four polarization vectors: the first ${\epsilon_{(0)}}^\mu$ coincides with a time-like vector  $n^\mu$ with $n^2=1$, then we consider two vectors ${\epsilon_{(\lambda')}}^\mu, \lambda'=1,2$ orthogonal to $P^\mu$  and $n^\mu$,  and finally the fourth one, ${\epsilon_{(3)}}^\mu=\frac {P^\mu - n^\mu  (P\cdot n)}{(P\cdot n)}$, 
orthogonal to the previous ones. These four vectors are orthonormal 
\be
 {\ep_{(\lambda)}}^\mu {\ep_{(\rho)}}^\nu\h_{\mu\nu}=\h_{(\lambda)(\rho)}.
 \ee
Using them we make a  canonical transformation  generated by 
 \be
 W(P,q,\bar x,u) =P_\mu \bar x^\mu+ u^{(\lambda)}\, {\ep_{(\lambda)}}^\mu\,q_\mu.
 \ee
 Here 
we take $n^\mu=(1,0,0,0)$ and the polarization vectors as
 \bea
 {\ep_{(\lambda)}}^\mu =\begin{pmatrix}
1&0&0&0\cr
0,&1 - \frac{{P_1}^2}
     {{|{\bf P}|}\,  ({|{\bf P}|} + P_3)   },&
   -  \frac{P_1\,P_2}
      {{|{\bf P}|}\,  ({|{\bf P}|} + P_3)   }
        ,&
   -  \frac{P_1}{{|{\bf P}|}}   \cr
 0,&-  \frac{P_1\,P_2}
      {{|{\bf P}|}\,  ({|{\bf P}|} + P_3)   }
       ,&
   1 - \frac{{P_2}^2}
     {{|{\bf P}|}\,  ({|{\bf P}|} + P_3)},&
   -  \frac{P_2}{{|{\bf P}|}}   \cr
 0,&\frac{P_1}{{|{\bf P}|}},&
   \frac{P_2}{{|{\bf P}|}},&\frac{P_3}{{|{\bf P}|}}\} \cr
   \end{pmatrix}.
   \eea
It follows
 \bea
x^\mu&=&\frac{\pa W}{\pa P_\mu}= \bar x^\mu +
u^{(\lambda)}\,\frac{\pa {\ep_{(\lambda)}}^\nu}{\pa P_\mu}\,q_\nu, 
\qquad
\ba P_\mu=\frac{\pa W}{\pa \bar x^\mu}=P_\mu,\quad
\nn\\
r^\mu&=&\frac{\pa W}{\pa q_\mu}=u^{(\lambda)}\, {\ep_{(\lambda)}}^\mu ,\qquad
v_{(\lambda)}=\frac{\pa W}{\pa u^{(\lambda)}}= {\ep_{(\lambda)}}^\mu\,q_\mu.
\eea

After quantization the commutation relations between the position and momentum operators
are
\be
[\bar x^\mu,P_\nu ]=-i{\D^\mu}_\nu,\quad \,\,\,\, [u^{(\lambda)},v_{(\rho)}]=-i
{\D^{\lambda}}_{\rho}
\ee 
and the annihilation $a_{(\lambda)}$ and creation operators
$a_{(\lambda)}^\dagger$ are given by
\be
a_{(\lambda)}=\frac 1 {\sqrt{2{\kappa}}}(v_{(\lambda)}- i {\kappa} u_{(\lambda)} ) ,
\qquad
a^\dagger_{(\lambda)}=\frac 1 {\sqrt{2{\kappa}}}(v_{(\lambda)}+ i {\kappa} u_{(\lambda)}    ) 
\ee
such that
\be  
[a_{(\lambda)},a^\dagger_{(\rho)}]=-g_{\lam\rho}.
\ee

 The constraints on the einbein sector give for sector 2
\be
\bar\pi_+|\psi_{phys}\rangle =0, \quad \bar\pi_-|\psi_{phys}\rangle =0,
\ee
which imply, in coordinate representation,  that the wave function does not depend on the einbein variables.
Instead, for the sector 3 we have
\be
\bar\pi_+|\psi_{phys}\rangle =0, \quad \bar a|\psi_{phys}\rangle =
(\bar e_-+i\bar\pi_-)|\psi_{phys}\rangle=0.
\ee
In this case   the wave function, in the coordinate representation, does not depend on $\bar e_+$ and is a gaussian function in the variable ${\bar e_-}$.

The constraints 
(\ref{relat})  become  operatorial conditions on the quantum states
\be
P^2\vert \psi_{phys}\rangle =0,\qquad 
(\sum_{\lambda'=1}^2 a^\dagger_{(\lambda')} a_{(\lambda')} -\beta) \vert \psi_{phys}\rangle =0.
\ee
Here $\beta$ is the parameter taking into account possible ambiguity in defining the 
quantum operator $a^\dagger_{(\lambda)} a_{(\lambda)}+ \frac {m_0^2}{2k}$. Note that we have included also the mass term in this ambiguity.
Finally the constraints on the transversality to the total momentum (\ref{trans})
are imposed \`a la Gupta Bleuler,
\be
(a_{(0)}-a_{(3)})\vert \psi_{phys}\rangle=0.
\label{transv2}
\ee
Therefore the construction of the Hilbert space can be performed by following the covariant quantization of the electromagnetic field.
The general solution of eq.(\ref{transv2}) can be written
as
\be
\vert \psi_{phys}\rangle =\vert \psi_T\rangle  \otimes \vert \phi_0\rangle  +\sum_{n\neq 0} c_n \vert \phi_n\rangle, 
\ee
where $n$ denotes the number of longitudinal $a_{(3)}$ and scalar $a_{(0)}$ oscillators  
and $\vert \psi_T\rangle $ is built with the transverse operators $a^\dagger_{(1)}$ and $a^\dagger_{(2)}$.
All the states $\vert \phi_n\rangle $ with $n\neq 0$ have zero norm and therefore at the end $\vert \psi_{phys}\rangle $  has positive definite norm.

Summing up only $\vert \psi_T\rangle $ contribute to the observables 
and the physical states exist for non-negative integers of $\beta$.  
If we choose an  integer $\beta=n\geq0$, the states have helicities ranging  
\be
n, n-2,\cdots, -n+2,-n.
\ee
For instance the states for $\beta=1$ are
\be
\frac 1 {\sqrt{2}}(a_{(1)}^\dagger\pm i a_{(2)}^\dagger)\vert p;0\rangle ,
\ee
where
\be
\vert p;0\rangle =\vert p\rangle \otimes \vert 0\rangle 
\ee
with $P_\mu \vert p\rangle =p_\mu \vert p\rangle $. They are 
corresponding to helicity $\pm 1$ states as can be checked from the form of the helicity operator,
\be
\Lambda=-i (a_{(1)}^\dagger a_{(2)} -a_{(2)}^\dagger a_{(1)}).
\label{helic}
\ee
 Indeed the Lorentz generators 
\be
M_{\mu\nu}=x_\mu p_\nu-x_\nu p_\mu+r_\mu q_\nu-r_\nu q_\mu
\label{angmomL}
\ee
take representation of massless little group  as 
\bea
M_{12}&=& \bar x_1 P_2-\bar x_2 P_1+\Lambda
\nn\\
M_{23}&=& \bar x_2 P_3-\bar x_3 P_2+\frac{P_1}{P_3+{|{\bf P}|}}\,\Lambda,
\quad 
M_{31}= \bar x_3 P_1-\bar x_1 P_3+\frac{P_2}{P_3+{|{\bf P}|}}\,\Lambda,
\nn\\
M_{01}&=&\bar x_0 P_1-\bar x_1 P_0-\frac{P_0P_2}{{|{\bf P}|}(P_3+{|{\bf P}|})}\,\Lambda,
\quad 
M_{02}=\bar x_0 P_2-\bar x_2 P_0+\frac{P_0P_1}{{|{\bf P}|}(P_3+{|{\bf P}|})}\,\Lambda,
\nn\\
M_{03}&=&\bar x_0 P_3-\bar x_3 P_0,
\eea
where $P_0=\pm |{\bf P}|$ corresponding to its signs. 

If we identify 
 $\beta=-2-\frac {m_0^2}k$, only $m_0^2<0$ can allow massless solutions for the physical states. The $P^2=0$ sector at the quantum level, and also at  classical level,  is empty for positive rest mass $m_0^2>0$. 
Non trivial massless sectors appear by  choosing the 
 tachyonic rest mass parameter $m_0^2<0$.  

\section{Conclusions and Outlook}

In this paper we have studied a model of two interacting relativistic particles via an harmonic potential at classical
and quantum levels. 
The model contains three parameters, the two rest masses of the particles and the frequency of the harmonic oscillator.
When the rest masses are different, the model has only the massive sector. At quantum level the spectrum in general contains a branch with ghosts.
Instead if the rest masses are equal, the model has the massive and massless sectors.
At quantum level the massive sector has a spectrum of increasing masses with higher internal spin.  The massless sector is non trivial at classical  
and quantum level when 
we consider tachyonic rest masses.
We have also given the two Noether gauge transformations  
in the massless sector of equal rest masses.
One of the transformations is the  worldline diffeomorphism. Instead for
the second gauge transformation, that has an open algebra, we have not yet found a clear
geometrical interpretation.

It is interesting to ask what happens when the distance among the particles is  constrained to be light-like.
This situation does not occur for the harmonic potential as we have seen. Light-like  
sectors with $r^2=0$, in addition to the massive sector, appear, 
 for example if we consider  a quartic potential $V({\r2})=k{(r^2)}^2$. 
Being more precise there are four sectors with the light-like constraint, 
among these there is one with the following set of constraints, 
\be
P^2=0,\quad q^2=0,\quad  r^2=0,\quad   (P\cdot r)=0,\quad  (P\cdot q)=0.
\ee
The same constraints appear in the rigid string model \cite{Casalbuoni:1974tq} \cite{Casalbuoni:1975mf}
\cite{Kamimura:2008ft}.
Therefore our model, for a suitable choice 
of the potential, has a sector that  describes the rigid motion of a string. 

Our model can be thought as a sort of effective theory describing possible interaction
 among two particles. It 
does not give any prescription to determine the potential. The form of the potential
 could be chosen from phenomenological reasons or 
in the best scenario could come from a  fundamental theory describing
 the higher energy degrees of freedom. 

We also think that the analysis of this model, presented here, could be an useful example to examine  other dynamical systems with sectors, 
like bigravity theories.

In a work in progress we will consider how we can construct the gauge transformations from the constraint structure.
We will also consider the gauge structure of the model by considering the solutions of the classical master equation.

\section*{Acknowledgments}
We acknowledge Jorge Alfaro, Luis Alvarez-Gaume, 
Max Banados, Carles Batlle, Eric Bergshoeff, Roberto Casalbuoni, 
Marc Henneaux,  Luca Lusanna,  Paul Townsend and Jorge Zanelli 
for comments and discussions. 

JG acknowledges the hospitality at the Universidad Cat\'olica de Santiago de Chile 
where this work has started and to CERN for the hospitality and partial financial support.
JG also acknowledges financial support from FPA 2010-20807, 2009 SGR502, CPAN, Consolider CSD 2007-0042
and Eplanet for the partial financial support for the  stay in Chile.

\end{document}